\documentclass{jaa}
\usepackage{natbib}
\usepackage{graphicx}
\usepackage{xcolor}

\begin{document}\sloppy

%%paper title
%%For line breaks \\ can be used within title
\title{Quasi-stationary sequences of hyper massive neutron stars with exotic equations of state}

%%author names are separated by comma (,)
%%use \and before the last author name
%%use a * along with the number separated by comma
%% for the  author for correspondence
%%\textsuperscript{number} is used for affiliation
%%\affilOne, \affilTwo etc., upto \affilTwentyfive is possible
%%Please note the first letter after \affil is capitalised in the command
%%

\author{Sanika Khadkikar\textsuperscript{1}, Chatrik Singh Mangat\textsuperscript{1} and Sarmistha Banik\textsuperscript{1}}
\affilOne{\textsuperscript{1} Department of Physics, Birla Institute of Technology and Science, Hyderabad}

%%escape two column mode for title, affiliation and abstract
%%by giving \twocolumn command as shown

\twocolumn[{

\maketitle

%%include \corres to print the corresponding author Email id
\corres{sarmistha.banik@hyderabad.bits-pilani.ac.in}

%%include \msinfo for
%%manuscript information such as
%%received, revised and accepted dates
%%
\msinfo{7 January 2022}{1 April 2022}

%%abstract
\begin{abstract}
In this work, we study the effect of differential rotation, finite temperature and strangeness on the quasi stationary sequences of hyper massive neutron stars (HMNS). We generate constant rest mass sequences of differentially rotating and uniformly rotating stars. The nucleonic matter relevant to the star interior is described within the framework of  relativistic mean field model with the DD2 parameter set. We also consider the  strange $\Lambda$ hyperons using the BHB$\Lambda\phi$ equation of state (EoS). Additionally, we probe the behaviour of neutron stars (NS) with these compositions at different temperatures. We report that the addition of hyperons to the EoS produces a significant boost to the spin-up phenomenon. Moreover, increasing the temperature can make the spin-up more robust. We also study the impact of strangeness and thermal effects on the T/W instability. Finally, we analyse equilibrium sequences of a NS following a stable transition from differential rotation to uniform rotation. The decrease in frequency relative to angular momentum loss during this transition, is significantly smaller for EoS containing hyperons, compared to nucleonic EoS. 
\end{abstract}

%%insert keywords separated by 3 hyphens using \keywords{words}

}]
%%close the twocolumn escape here

%%include \doinum{number}for the DOI number in the header
%%include \volnum{number} for the volume number in the header
%%include \year{yyyy} for  year of publication in the header
%%include \pgrange{num--num} page range of article in the header
%%include \artcitid{num} for the article citation id
%%include \lp to print last page of the article
%%include \setcounter{page}{pagenum} for the exact starting page of the article

\doinum{12.3456/s78910-011-012-3}
\artcitid{\#\#\#\#}
\volnum{000}
\year{0000}
\pgrange{1--11}
\setcounter{page}{1}
\lp{11}

\section{Introduction}
Constraining the Equation of State (EoS) of Neutron Stars (NS) has been one of the biggest unsolved problems for the Nuclear Astrophysics. The dense core
is inaccessible to any sort of observation, hence its composition is unknown to us. There has been a lot of speculation on the composition of the NS interior, based on our knowledge of terrestrial experiments such as
nuclear physics and heavy ion collisions. Various theoretical models and EoSs for the NS are constructed based on these clues, though the model parameters are extrapolated to asymmetric nuclear matter, 
well above the nuclear matter density. These microscopic EoSs are then used to generate the static structure of the stars,  which  on the other hand can be compared with the astrophysical observables - mass and radius.

With the advancement of observational techniques, we continue to get increasingly accurate data to constrain the properties of NS. In the recent past the Neutron Star Interior Composition Explorer (NICER) could determine the  mass and radius of PSR J0030+0451 \citep{Raaijmakers2019} simultaneously for the first time. Two different analyses of NICER data estimate the radii of $13.02^{+1.24}_{-1.06}$ km for a 1.44$^{+0.15}_{-0.14}M_{sol}$ pulsar \citep{Miller2019} and 12.71$^{+1.14}_{-1.19}$ km for a 1.34$^{+0.15}_{-0.16}M_{sol}$ pulsar \citep{Riley2019} respectively. The BNS merger GW170817 event estimates the radius of 1.4$M_{sol}$ neutron star in the range $11.9^{+1.4}_{-1.4}$
km \citep{Abbott2018} from its tidal
%\bibitem{bpa} B. P. Abbott et al., Phys. Rev. Lett. \textbf{121} (2018) 161101.
deformability value of  
($70 \leq {\Lambda_{1.4}} \leq 580$). Mass-observations of galactic  pulsars PSR 0348+0432 of $2.01^{+0.04}_{-0.04} M_{sol}$ and %\textcolor{red}{PSR 0740+6620 of $2.14^{+0.10}_{-0.09} M_{sol}$  \citep{Antoniadis2013, Cromartie2020}} 
PSR 0740+6620 of $2.08^{+0.07}_{-0.07} M_{sol}$  \citep{Fonseca2021} have made the soft EoSs redundant.
%\textcolor{red}{These new observations prompt us to explore and tweak the theoretical framework which describes these compact objects.} 
These new observations help us constrain EoS parameters and improve our understanding of the dynamical processes occuring in these compact objects.

The maximum possible mass of a non-rotating (static) NS depends heavily on its EoS. But, all physical NS we expect to observe will be rotating at high frequencies, since the angular momentum of the progenitor is transferred to the remnant core in a Type II Supernova. This uniform rotation can allow a NS to support around 8-20\% more mass than a static NS \citep{Khadkikar2021}. 
%For these rotating NS, the Kepler mass-shedding limit determines the maximum rotation rate of a NS, which in turn limits the maximum mass. 
A rotating NS which is too massive to be supported without rotation is called a supramassive NS, while a star with a stable static configuration is called a normal NS.
Moreover, the process of differential rotation (i.e. the core rotating at a higher frequency than the envelope) can significantly increase the limit on the maximum mass of a NS \citep{Morrison2004, apj2020}. These high mass NS supported by differential rotation are also known as Hyper Massive Neutron Stars (HMNS). Differential rotation is also speculated to stabilize Proto Neutron Stars (PNS) during a supernova event and the high mass remnants created by NS-NS merger events.
%which have masses too large to be supported by uniform rotation ($M \geq 2.9  M_{sol}$) \citep{Baumgarte2000}}.

We consider only the most realistic solutions of differentially rotating stars that belong to the spheroidal  class (Type `A') \citep{Ansorg2009}, which always have  a mass-shedding limit. When the degree of differential rotation is large, the maximum density moves away from the rotation axis, resulting in a toroidal configuration \citep{Galeazzi2012}.
% When the degree of differential rotation is large the maximum density may move away from the rotation axis. When the value of the axis ratio reaches zero, the star develops a hole on the rotation axis. The equilibrium sequence may be continued beyond this point, but we are not interested in this toroidal shape configuration as a model of a neutron star.
%F. Galeazzi, S. Yoshida, and Y. Eriguchi,A\&A 541, A156 (2012)
But, the spheroidal class sequence of relativistic solutions  is not connected to the regime of configurations with toroidal shape. Instead, the latter may lead either to a mass-shedding limit or to a limit of infinite central pressure \citep{Ansorg2009}.
We investigate the species of differentially rotating neutron stars in greater depth, wherein we follow the quasi-static evolution of a differentially rotating NS down to its static counterpart, if any. Hence we restrict our calculations to the spheroidal class.

%\textbf{EoS \& exotic matter}
A NS-NS merger can result in a HMNS remnant with densities up to 3-5 times normal nuclear matter density ($n_0$) and temperatures $\sim 50$ MeV  \citep{Perego2019,Bernuzzi2020}.
The high density HMNS core is believed to contain a few exotic particles, such as hyperons, quarks and/or antikaon condensate, making the EoS softer \citep{apj2020}. A strong quark-hadron phase transition may be detected in the inspiral as well as in post-merger phase \citep{Chatziioannou2020, Bauswein2019}. Appearance of strangeness is more feasible in post-merger remnants due to their high central density and temperature conditions. Once the HMNS is formed, large shocks will increase the temperature, and to account for this additional heating, a finite-temperature  EOS is required \citep{Hanauske2017}. 

Realistic EoS can support two kinds of equilibrium sequences in an isolated NS - normal and supramassive. Normal sequences, which contain a stable non-rotating profile, are considered stable to quasi-radial perturbations. On the other hand, all supramassive stars have a rest mass which exceeds the maximum mass of a stable non rotating NS. It has been shown by various authors that the presence of hyperons, $K^-$ condensates and quark matter in an  isolated NS  can spin-up its uniform rotation \citep{Zdunik2004, Banik2004, Spyrou2002, Glendenning2000}. The spin-up is accompanied by a loss in angular momentum due to changes in the moment of inertia as a supramassive star evolves towards the radial instability point. Beyond this point, both angular momentum and rotational frequency start increasing, wherein the star crosses into unstable sequence territory. This behaviour is also known as the back-bending phenomenon for rotating NS.

When dealing with NS evolution, it is very important to look at the stability with respect to the axi-symmetric perturbations \citep{Zdunik2004}.
%Zdunik, J. L., Haensel, P., Gourgoulhon, E., & Bejger, M. 2004, A&A, 416, 1013
An important measure of rotation is the T/W parameter i.e. the ratio of rotational kinetic energy to gravitational binding energy, introduced
by \cite{Komatsu1989}.  The axial symmetry of a rotating NS is spontaneously broken if T/W exceeds a critical value; this marks the onset of secular (i.e., viscosity driven) instability.  For  uniformly
rotating, incompressible ellipsoids, the onset of secular and gravitational wave-emission instabilities
arise at the lowest order for l = 2 modes
and $m = \pm 2$ perturbations in Newtonian  and
general relativistic systems.
A limiting value  of the ratio T/W = 0.128 was found for a differentially-rotating NS with a polytropic EoS \citep{Galeazzi2012}.
A differentially rotating NS is reported to be dynamically unstable for T/W  \textgreater{ 0.24} \citep{Shibata2000, Baiotti2007}.
Relatively low values of T/W $\sim 0.01$ were  calculated to establish the instabilities in NS merger remnants \citep{Shibata2003, Passamonti2020}.
The NS remnant can either collapse into a black hole or settle as an axisymmetric  equilibrium configuration. While our work uses the barotropic assumption for the EoS where the thermodynamic properties of the NS are a function of the pressure, there have been studies of NS with non-barotropic profiles \citep{Camelio2019}. The HMNS is also believed to be surrounded by an accreted disk of mass $\sim 0.01 M_{sol}$ which may affect the long-term stability of the remnant \citep{Bernuzzi2020, Xie2020}. However, this is beyond the scope of this paper, and we leave it for a future study.
%Bernuzzi https://arxiv.org/pdf/2004.06419.pdf
%Xie https://arxiv.org/pdf/2005.13696.pdf

%They considered  possible outcomes of the NS-NS merger. A HMNS may form and survive for 10-100 ms  if the nascent NS mass ($M_p$) is greater than the maximum mass that may be supported by uniform rotation; the NS may instantly collapse to a black hole also \citep{Ravi2014}. 
%Vikram Ravi & Paul Lasky, Mon. Not. R. Astron. Soc. 000, 1–8 (2014)

The outline of the paper is as follows. In Section \ref{sec:method}, we present the formalism, namely the rotation law for differential rotation, and a brief description of the  different relativistic EoSs employed in this investigation. We then discuss the numerical scheme employed to obtain the macroscopic NS structure. 
In Section \ref{sec:results}, we report the main results of this study.
We provide the details of the method used to follow the quasi-static equilibrium sequences for various differential rotation parameters and EoSs. In Section \ref{sec:sum}, we summarise the main findings of this work, its limitations, and the scope for further study.
%\subsection{Subsection heading}
%Subsection text here.
%\subsubsection{Subsubsection heading.} Subsubsection text goes here (Radhakrishnan {\em et al.} 1980).
\section{Equation of state and Numerical Scheme of rotating NS}
\label{sec:method}
The EoS we work with are realistic, applicable for a charge neutral, $\beta$-equilibrated NS matter. We consider a NS with (i) pure nucleonic core (DD2 EoS) consisting of neutrons, protons and electrons \citep{Hempel2010},
%NPA837 (2010) 210
(ii) core with additional $\Lambda$-hyperons  (BHB$\Lambda\phi$ EoS) \citep{BHB}.  Two different
variants of the EoSs are used, one corresponding to cold NS (T=0) and the other having constant entropy per particle (s=2 in units of Boltzmann constant $k_B$).  Henceforth we consider $k_B=1$ throughout the work.
The role of finite temperature on the NS merger remnant  has been explored for s=0 and s=2 cases \citep{apj2020, Soma2020}. We use these values to represent cold and hot stars since studies conclude that the entropy per baryon is $s<2$ at the core of the remnant just after the merger whereas the bulk of the remnant outside the unshocked core has the entropy per baryon which is a few times $k_B$ \citep{Perego2019}. Although there is a large spread in temperature and entropy initially, the conditions become homogeneous at a later time.

The hadronic part of both the EoS is computed in the framework of relativistic mean field model (RMF) \citep{Serot1986, Walecka1995, Hempel2010, Glendenning2000} where the constituent baryons interact via $\sigma$, $\omega$ and $\rho$ mesons. Thebaryon-meson couplings are made density-dependent to take care of the high density behaviour of the dense matter, whose properties are not known. However, the model parameters are fitted to the properties of finite nuclei to obtain the bulk properties of nuclear matter at normal nuclear matter density ($n_0$). The saturation properties of symmetric
nuclear matter 
%for DD2 and BHB$\Lambda\phi$ EoSs 
are $n_0 =$ 0.149065 fm$^{-3}$, binding energy per nucleon 16.02 MeV, incompressibility 242.7 MeV \citep{Typel2010}. The models agree with the values of  symmetry energy 31.67 MeV and its slope coefficient
55.03 MeV, set by experiments. These models can also reproduce the finite nuclei properties. Further, the pure neutron matter calculation with the DD2 parameter set is in agreement with that of the chiral effective theory \citep{Marques2017}. The inhomogeneous nuclear matter at sub-saturation density is composed of light and heavy nuclei along with unbound nucleons and is described within the extended Nuclear Statistical Equilibrium (NSE) Model of \cite{Hempel2010}. This is merged with the high density part of our EoS by minimisation of free energy, the thermodynamic consistency is ensured in the process \citep{BHB}. The EoSs at sub-saturation density as well as high density, are computed using the same RMF nucleon-nucleon interaction and can be considered unified.   Moreover, the observational constraint of 2$M_{sol}$ \citep{Demorest2010, Antoniadis2013} on their mass is satisfied, even in the presence of strange $\Lambda$ particles.
\begin{figure*}[!h]
    \includegraphics[width=1.0\textwidth]{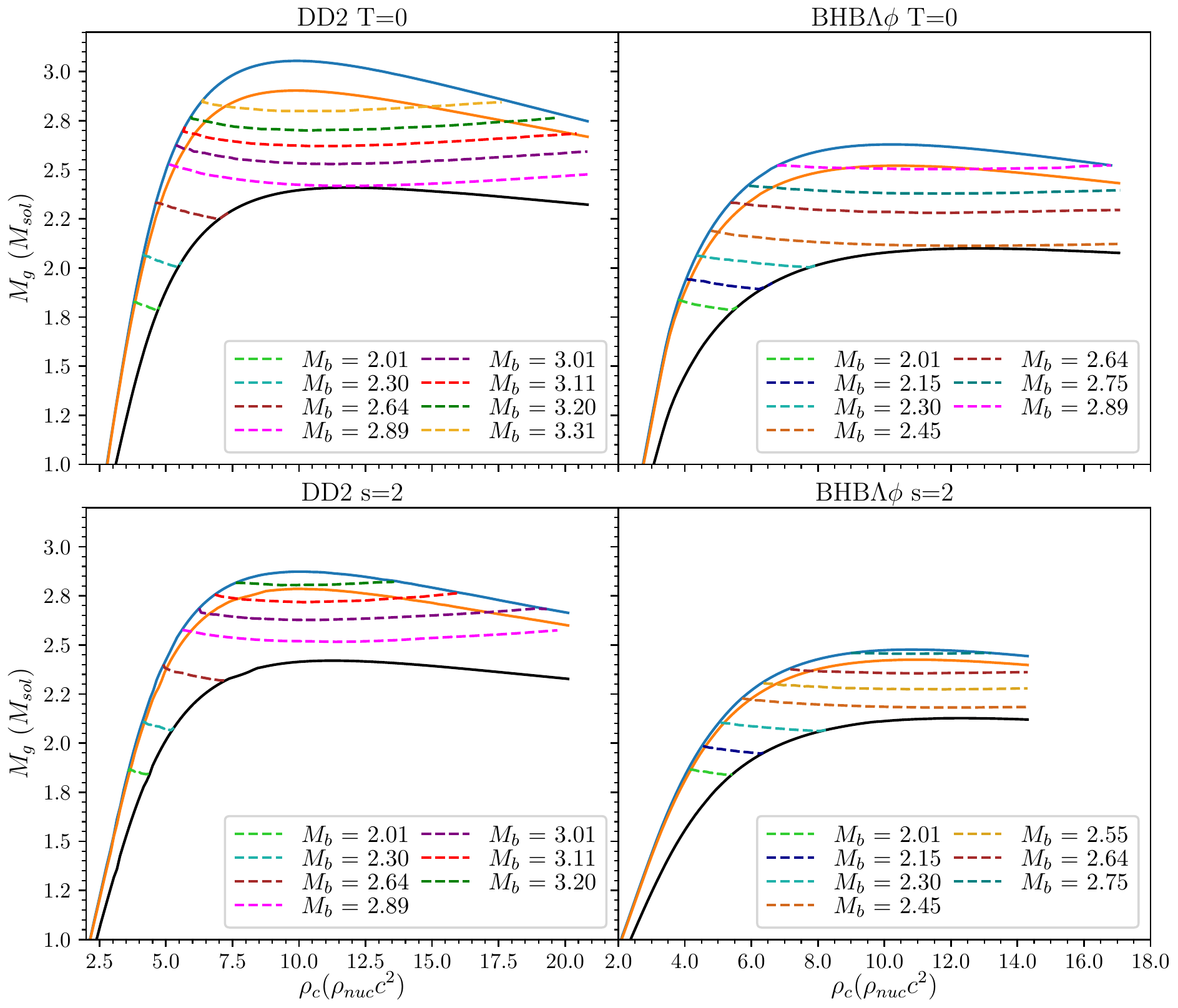}
    \caption{Gravitational mass vs central energy density of NS with DD2 and BHB$\Lambda \phi$ EoS for T=0 and s=2 configurations. The solid lines represent static (black), uniformly rotating (orange) and differentially rotating with $a$ = 0.2} (blue) stars at Kepler frequencies. The dashed lines are constant rest mass (in $M_{sol}$) sequences (color online).
\label{Fig:Mass_rho}
\end{figure*}
$\Lambda$'s are populated at the cost of  neutrons, when their chemical potentials attain equilibrium. We use the BHB $\Lambda \phi$ EoS for strange matter, that is an extension of DD2 EoS. The additional $\phi$ mesons take care of the $\Lambda-\Lambda$ interaction.
%equilibrium condition $\mu_n = \mu_\Lambda$ is met. 
The density-dependent $\Lambda$-vector meson hyperon vertices are obtained
from the SU(6) symmetry of the quark model  \citep{Dover1984,Schaffner1996}. 
%$g_{\omega \Lambda} = \frac{2}{3} g_{\omega N}, \quad g_{\rho \Lambda} = 0, \quad g_{\phi \Lambda} = -\frac{\sqrt 2}{3} g_{\omega N}$.
On the other hand, the $\gamma$-ray spectroscopy for $\Lambda$-hypernuclei indicates a potential depth of around 30 MeV. The $\Lambda$ hyperon - scalar meson couplings are obtained from the potential of  $\Lambda$ hyperons in nuclear normal matter and the vector meson couplings.
The ratio of $g_{\sigma \Lambda}$ to $g_{\sigma N}$ is 0.62008 \citep{BHB}.
%We compute rest mast sequences for T=0 and s=2 variants. 
However, the nature of hyperon-nucleon interaction cannot be ascertained without ambiguity because of limited  scattering data and measured properties of hyper-nuclei, for other hyperons like $\Sigma$ and $\Xi$. Hence, we keep our calculations restricted to the lightest hyperon of the octet.

We use codes based on the LORENE \citep{lereh2016} library to study rapidly rotating NS. There are already existing numerical schemes that compute equilibrium solutions of uniform and differentially rotating cold NSs. It can also deal with isentropic EoS \citep{Goussard1997, Goussard1998}. The algorithm to construct rotating equilibrium models numerically in a
self-consistent-field method is explicitly described in \cite{Bonazzola1993, apj2020}.
%generate the neutron star rest mass sequences for our work.
We  employ the simple KEH \citep{Komatsu1989} or j-constant rotation law defined by the velocity profile to compute various properties of the differentially rotating HMNS:

$$
F(\Omega)=R_{0}^{2}\left(\Omega_{c}-\Omega\right)
$$

where $\Omega_{c}$ is the central angular frequency. The angular velocity decreases monotonically from core to the surface of the star and the degree of differential rotation is encapsulated in the parameter $a=R_e/R_0$ where $R_e$ is the equatorial radius of the NS. Our results apply particularly to the class of solutions of ‘type A’, as classified by \cite{Ansorg2009}. These solutions possess a mass-shedding limit, and their rest-mass density is the maximum at the stellar centre.  Higher values of $a$ will lead to higher angular momenta $J$, which will shift the mass-shedding limit to higher masses \citep{Weih2018}, and might move the maximum density away from the rotational axis. Moreover, an overall mass-shedding limit may not exist for higher degrees of differential rotation. Hence, throughout this work, we have taken $a=0.2$ unless specified otherwise. We observed that the numerical fluctuations in the differential Kepler frequency curves could be reduced by giving a high frequency as input with a $10^{-6}$ order convergence threshold. Incidentally, the differential rotation law with the specific angular momentum is not constant in general, but only in a particular limit and in Newtonian gravity \citep{Camelio2021}. A number of binary NS merger simulations suggest the angular velocity profile of the HMNS is characterized by a slowly rotating core and an envelope that rotates at frequencies scaling like $r^{-3/2}$  \citep{Hanauske2017} unlike the j-constant law.
Uryu et al. (2017) formulated 
new  models with 3 and 4 parameters for differential rotation of compact stars \citep{Uryu2017}. 
Several authors have worked on more realistic  rotation profiles \citep{Bozzola2018, Iosif2021, Camelio2021} to construct equilibrium models of differentially rotating BNS remnants. Iosif and Stergioulas have compared the Uryu differential law and KEH law and concluded  that the gravitational mass does not have any difference for type A, B, C sequences of  \citep{Ansorg2009}, but the choice of rotation law has larger effect on  the radius \citep{Iosif2021}. Passamonti and Andersson  studied the impact of differential rotation on the 
low T/W instability using the 3-parameter rotation law of Uryu and highlighted the similarity with the widely used j-constant law \citep{Passamonti2020}. Hence, we stick to the commonly used j-constant law in this work to first observe the effects of finite temperature and strangeness on differentially rotating NS. We will explore the effect of modified rotation laws in this regime in future work.
%%Use table environment for a table in one column
\begin{table*}[!h]
    \tabularfont
    \centering
    \begin{tabular}{|c|c|c|c|c|c|c|c|}
    \hline
    \multicolumn{8}{|c|}{ [ DD2, $T=0$ ]} \\
    \hline
          Configuration & $M_{b}(M_{sol})$ & $M_{g}(M_{sol})$ & Radius($km$) & $\rho_c^{max}(\rho_{nuc}c^2)$ & $f_c(Hz)$ & $J(G{M_{sol}^2}/c)$ & T/W \\
         \hline
         Differential Kepler & 3.60 & 3.05 & 11.32 & 9.99 & 1783.25 & 7.16 & 0.16 \\
         Uniform Kepler & 3.43 & 2.90 & 10.83 & 9.82 & 1535.34 & 5.92 & 0.14 \\
         Static & 2.88 & 2.41 & 8.01 & 11.96 & 0.0 & 0.0 & 0.0 \\
    \hline
    
    \hline
    \multicolumn{8}{|c|}{ [ DD2, s=2 ]} \\
    \hline
         Configuration & $M_{b}(M_{sol})$ & $M_{g}(M_{sol})$ & Radius($km$) & $\rho_c^{max}(\rho_{nuc}c^2)$ & $f_c(Hz)$ & $J(G{M_{sol}^2}/c)$ & T/W  \\
         
         \hline
         Differential Kepler & 3.28 & 2.87 & 11.95 & 10.24 & 1537.87 & 5.55 & 0.12 \\
         Uniform Kepler & 3.18 & 2.78 & 11.65 & 9.81 & 1343.29 & 4.81 & 0.10 \\
         Static & 2.79 & 2.42 & 8.65 & 11.18 & 0.0 & 0.0 & 0.0 \\
    \hline
    
    \hline
    \multicolumn{8}{|c|}{ [ BHB$\Lambda\phi$, $T=0$ ]} \\
    \hline
         Configuration & $M_{b}(M_{sol})$ & $M_{g}(M_{sol})$ & Radius($km$) & $\rho_c^{max}(\rho_{nuc}c^2)$ & $f_c(Hz)$ & $J(G{M_{sol}^2}/c)$ & T/W  \\
         \hline
         Differential Kepler & 3.04 & 2.63 & 11.41 & 10.37 & 1573.40 & 4.98 & 0.14 \\
         Uniform Kepler & 2.91 & 2.52 & 10.92 & 10.48 & 1422.22 & 4.25 & 0.12 \\
         Static & 2.43 & 2.10 & 7.84 & 12.58 & 0.0 & 0.0 & 0.0 \\
    \hline
    
    \hline
    \multicolumn{8}{|c|}{ [ BHB$\Lambda\phi$, s=2  ]} \\
    \hline
         Configuration & $M_{b}(M_{sol})$ & $M_{g}(M_{sol})$ & Radius($km$) & $\rho_c^{max}(\rho_{nuc}c^2)$ & $f_c(Hz)$ & $J(G{M_{sol}^2}/c)$ & T/W  \\
         \hline
         Differential Kepler & 2.77 & 2.48 & 11.98 & 10.89 & 1384.66 & 3.78 & 0.10 \\
         Uniform Kepler & 2.71 & 2.42 & 11.63 & 10.90 & 1265.43 & 3.39 & 0.09 \\
         Static & 2.39 & 2.13 & 8.49 & 12.23 & 0.0 & 0.0 & 0.0 \\
    \hline
    \end{tabular}
    \caption{Properties of the maximum mass NS spinning at mass-shedding limits, rotating differentially with $a$ = 0.2} and uniformly. The static case corresponds to properties of the maximum mass non-rotating NS for each EoS. 
    \label{tab:max_mass}
\end{table*}
\section{Results}\label{sec:results}
We now present the results of our investigation. We compare the properties of NS generated using the DD2 EoS and BHB$\Lambda\phi$ EoS at zero temperature and finite temperature (s=2). We also explore the effect of varying degree of differential rotation on the evolution of the NS.
In Fig. \ref{Fig:Mass_rho}, we plot the gravitational mass (in solar masses) versus central energy density  (in units of $\rho_{nuc} c^2$, where $\rho_{nuc}=1.66 \times 10^{17} kg/m^3$). The top(bottom) panels are for cold(hot) stars, whereas the left(right) panels are for DD2(BHB$\Lambda \phi$) EoS. The equilibrium sequences for non-rotating (black solid curve) and mass-shedding or ``Kepler" limits of uniformly rotating NSs (orange solid curve) and differentially rotating NSs with $a=0.2$ (blue  solid curve) are plotted. We see that differential rotation allows a NS to support more mass than a non-rotating or a uniformly rotating NS. 
%Here in Fig.\ref{Fig:mgDD2s2} we compare the NS with DD2(BHB$\Lambda\phi$) EoS on the left(right) panel. 

When comparing stars with similar compositions, a hot static star, owing to thermal pressure, can support more mass than the cold one. However, we observe a different trend for both uniform and differentially rotating stars. The Kepler frequency (rotation rate corresponding to the mass-shedding limit) for hot stars is lower than that of cold stars. This is because the additional thermal energy in a hot star compensates for a part of the rotational energy, making the hot star start shedding mass at a lower rotation rate than a cold star. Hence, the maximum mass corresponding to a cold star rotating at its Kepler frequency is found to be more massive than that of a hot star. This is evident from the properties listed in Table \ref{tab:max_mass}.
\begin{figure*}[!h]
    \includegraphics[width=1.0\textwidth]{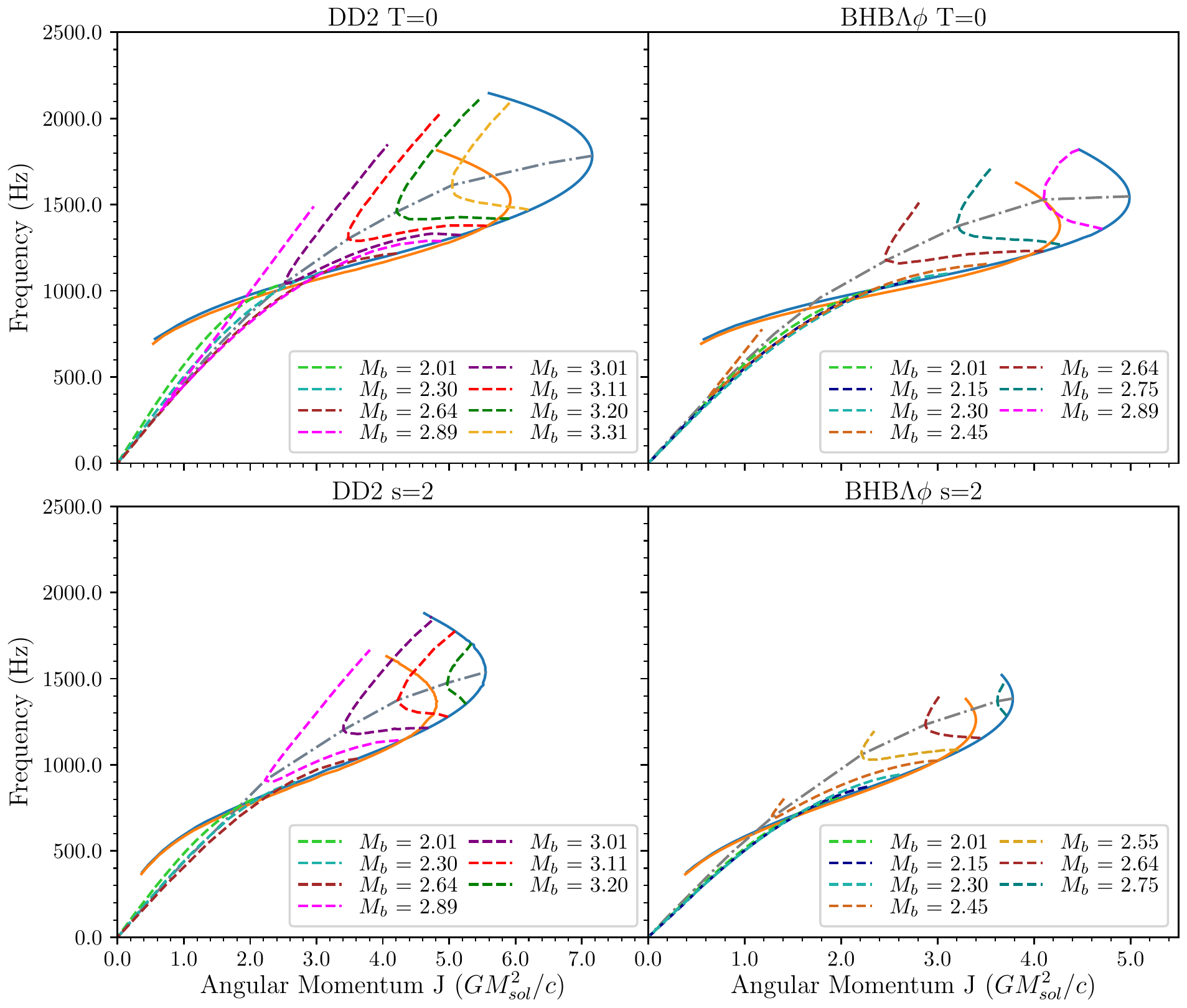}
    \caption{Central rotation frequency vs angular momentum of NS with DD2 and BHB$\Lambda \phi$ EoS for T=0 and s=2 configurations. The solid lines represent uniformly rotating (orange) and differentially rotating with $a$ = 0.2} (blue) stars at Kepler frequencies. The gray dash-dotted line indicates the quasi-stationary stability limit of the dashed constant-rest mass sequences (color online).
\label{Fig:Freq_j}
\end{figure*}
\begin{figure*}[h]
    \includegraphics[width=1.0\textwidth]{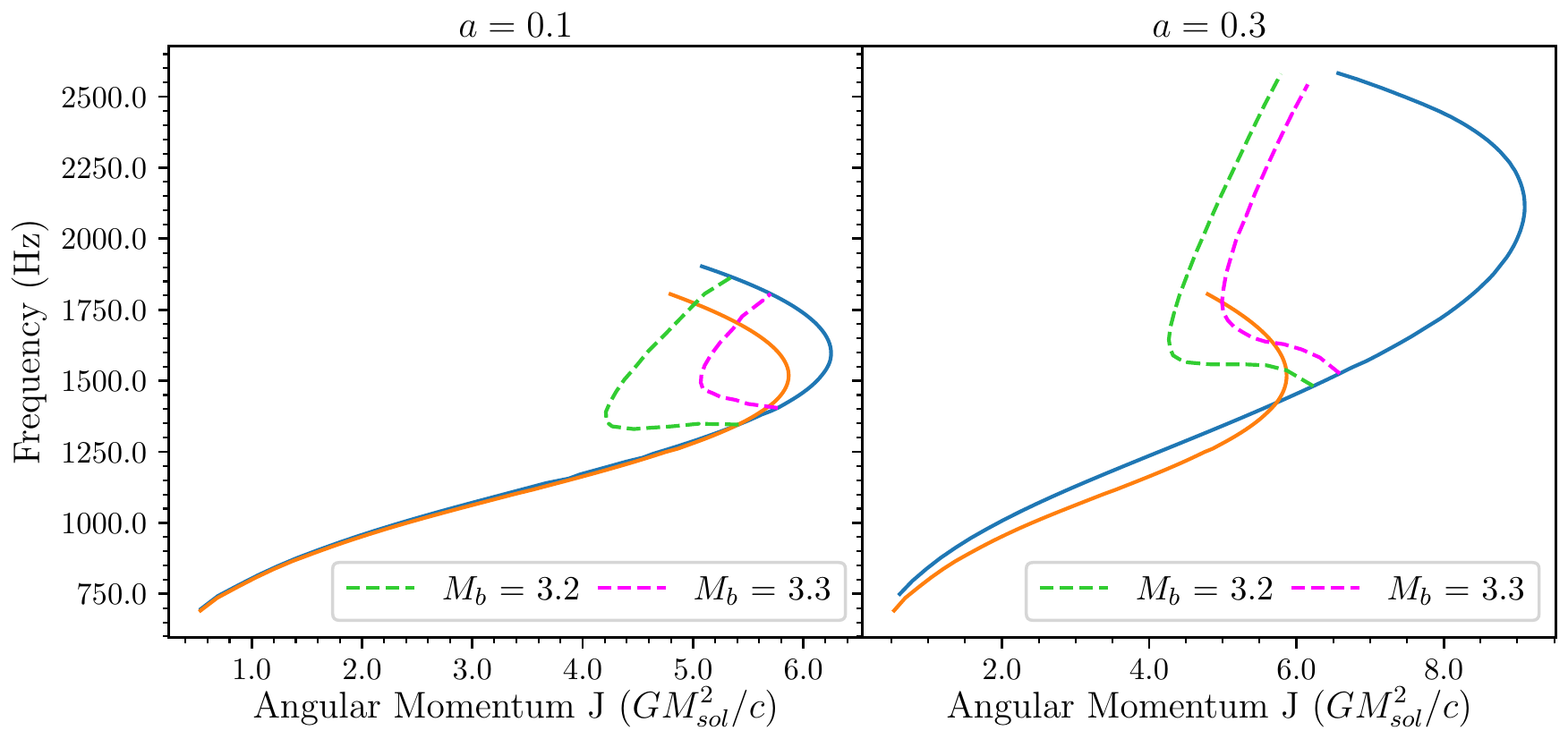}
    \caption{Rotational frequency as a function angular momentum for differentially rotating NS with the DD2 EoS at T=0 for different degrees of differential rotation $a$. The solid curves represent NS rotating at the mass-shedding limit uniformly (orange) and differentially (blue). The dashed curves represent constant rest mass sequences of differentially rotating NS (color online).}
\label{Fig:Freq_j_diffa}
\end{figure*}
\begin{figure*}[h]
    \includegraphics[width=1.0\textwidth]{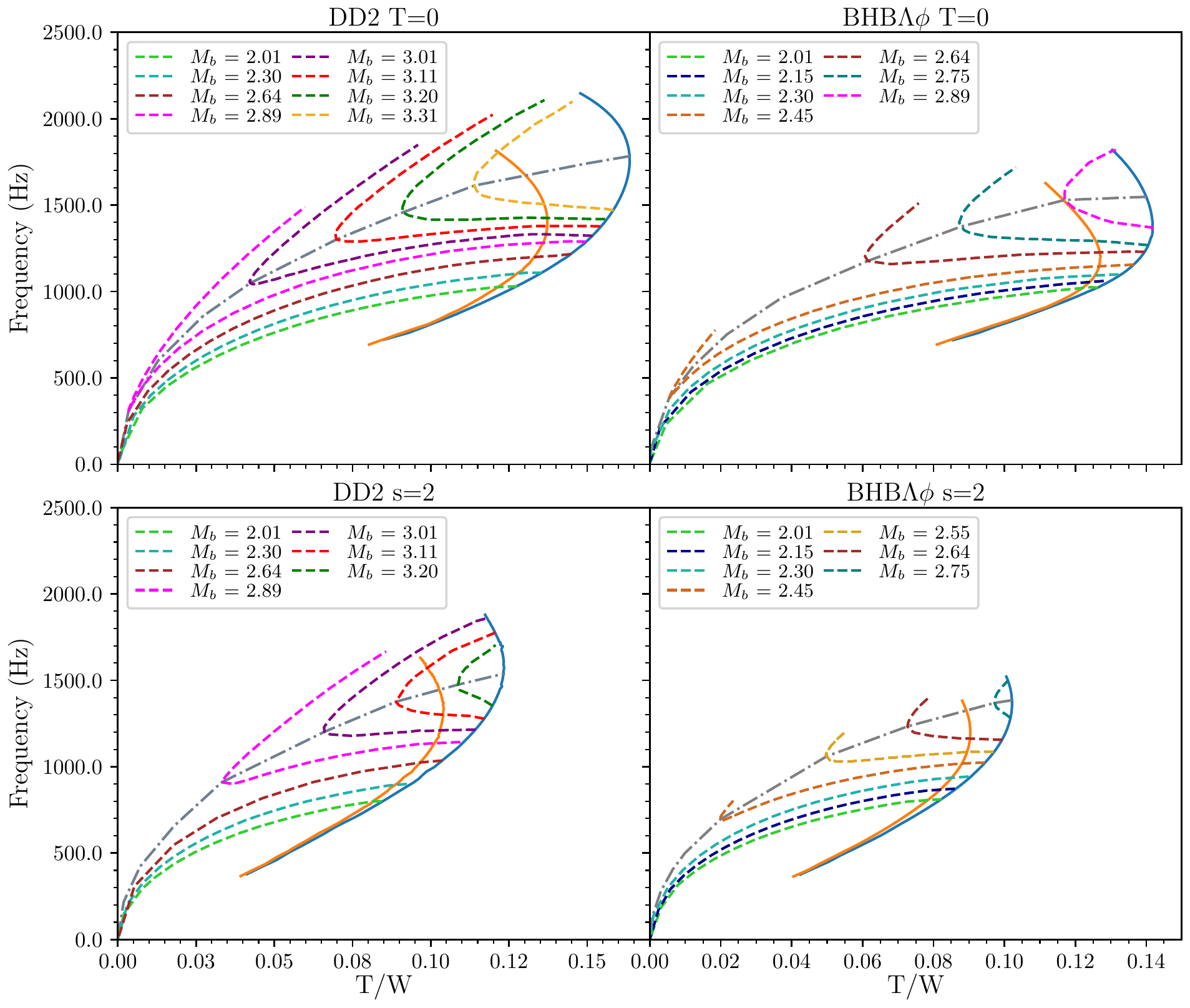}
    \caption{Central rotation frequency vs T/W of NS with DD2 and BHB$\Lambda \phi$ EoS for T=0 and s=2 configurations. The format of the graph is the same as Figure \ref{Fig:Freq_j}, with the dash-dotted line indicating the quasi-stationary stability limit (color online).}
\label{Fig:Freq_tw}
\end{figure*}
\begin{figure*}[!h]
    \includegraphics[width=1.0\textwidth]{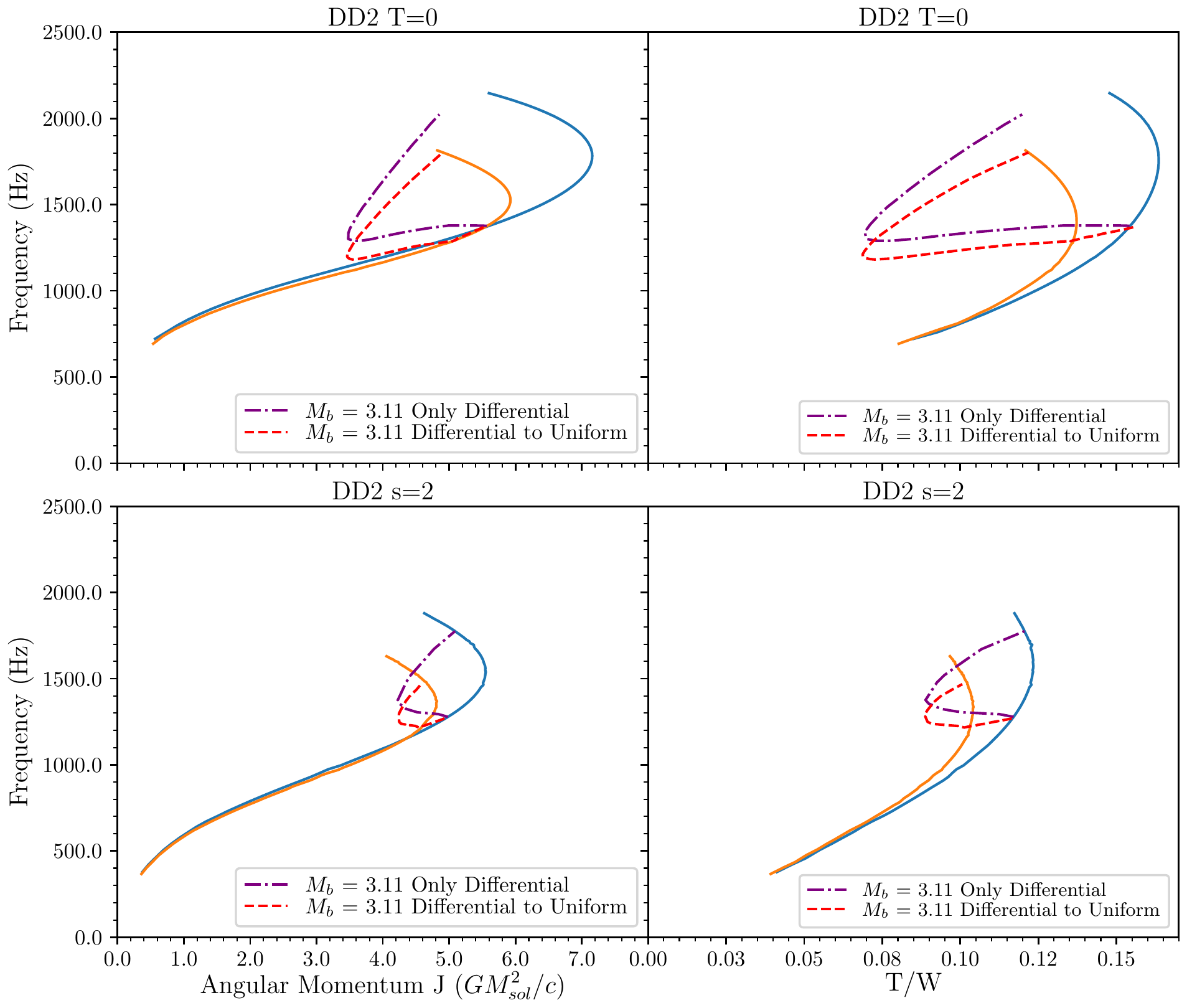}
    \caption{A comparison of spin-up in supramassive stars with $M_b= 3.11 M_{sol}$ for cold (top) and hot (bottom) NS with DD2 EoS. The stars represented by the purple dash-dotted curves rotate differentially with $a$ = 0.2} throughout their evolution. The red dashed curves represent  stars where differential rotation ($a$ = 0.2) decays to uniform rotation ($a$ = 0.0) as the star evolves. We also plot the Kepler sequences for uniform rotation (orange) and differential rotation (blue) for reference.
\label{Fig:Freq_comp}
\end{figure*}
Comparing the left and right panels of Fig. \ref{Fig:Mass_rho} reveals that the NS can support a smaller maximum mass in the presence of hyperons, other qualitative features remaining same as DD2 EoS. This reduction in the NS mass is consistent with the softening of the EoS due to the addition of hyperons.

The maximum mass $M_g^{max}$ allowed for each case of Fig. \ref{Fig:Mass_rho} is tabulated along with the central density $\rho_c^{max}$ in Table \ref{tab:max_mass}. The static NS masses are well within the observational constraint \citep{Antoniadis2013, Cromartie2020}. 
%\textcolor{green}{do we write the constraint values here?} 
We also observe that the relative increase in $M_g^{max}$ due to differential rotation is higher for cold stars than hot stars, for both the compositions.
% Rel increase = [diff - unif(stat)] / unif(stat)
% DD2 s0 DD2 s2 BHB s0 BHB s2 (unif stat)
%0.052 0.268 | 0.031 0.187 | 0.043 0.252 | 0.021 0.164 Mg
%0.050 0.250 | 0.029 0.175 | 0.046 0.249 | 0.022 0.160 Mb
The condition, $\frac {d M_g}{d \rho_c}<0$, marks the  onset of the secular instability for isentropic (or isothermal), rigidly rotating \citep{Goussard1997, Marques2017}  as well as for differentially rotating stars \citep{Bozzola2018, Weih2018} . The configurations with $\rho_c<\rho_c^{max}$ are stable
whereas those with $\rho_c>\rho_c^{max}$ with respect to small radial perturbations.   

We then plot constant rest mass sequences for differentially rotating neutron stars. Each sequence represents an isolated NS of fixed rest mass, that  rotates differentially and traces a curve of decreasing $M_g$ initially. 
The lower mass stars slowly lose energy  
and eventually meet the static curve when they stop rotating, these are called normal sequences. On the other hand, the higher mass sequences, also called supramassive sequences, never meet the static curve. The higher mass sequences trace a curve of decreasing mass to reach a minimum $M_g$ before rising again to meet the differential Kepler curve once again. The turn around point in these sequences is called the quasi stationary stability limit. The minimum $M_g$ configurations in these sequences correspond to the quasi stationary stability limit. These supramassive NS have masses larger than the maximum mass non-rotating NS. The behaviour is similar to that of uniformly rotating stars in \cite{Cook1994}.   
 For DD2 EoS, a 2.89 $M_{sol}$ baryon mass sequence barely touches the static curve for the T=0 case, whereas it does not have any static counterpart for the s=2 case. For BHB$\Lambda \phi$ EoS, this happens at $M_b \sim 2.45 M_{sol}$. Importantly, we note that the constant rest mass sequences, when plotted in Fig.\ref{Fig:Mass_rho}, do not vary much for (i) stars that rotate differentially for a brief period and then continue to rotate uniformly and (ii) stars that rotate differentially throughout their evolution.

%on a star consisting of nucleons only. We plot the gravitational mass versus central energy density, clubbing two values of differential rotation parameter a=0.1 and 0.3, 
%the figures for the same quantities 
%for both T=0 and s=2 DD2 EoS. 
%more the degree of differential rotation, more is the maximum mass.

In Fig. \ref{Fig:Freq_j}, we plot the rotational frequency against angular momentum to investigate the spin-up phenomenon. The solid curves mark the mass-shedding limits; we use blue for the differentially rotating stars with $a$ = 0.2 and orange for the uniformly rotating ones. The dashed lines are constant baryon mass sequences. Along normal sequences, the NS slows down as it evolves by radiating energy and angular momentum to move towards the origin. However, along supramassive sequences, the NS is observed to exhibit spin-up at some point in its evolution. This can be seen as back-bending, where the angular momentum decreases with increasing frequency for a short time, for all the cases considered here. The spin-up or back-bending rate is visibly more prominent for higher entropy stars. If we compare the massive NS with DD2 EoS of a particular baryon mass ($3.2M_{sol}$ for example), a star with entropy s=2 is observed to spin-up by $120Hz$, which is more than a cold star which spins up by $40 Hz$. Inversely, the angular momentum loss for the hot star is $0.28\text{ } GM^2_{sol}/c$, which is smaller compared to $0.49 \text{ } GM^2_{sol}/c$ for the cold star. The gray dashed-dotted line represents the quasi-stationary stability line, above which the supramassive sequences are unstable. We find that the quasi-stationary line can be fit with a polynomial function of the form 
$$
\dfrac{f_c}{f_c^\text{M}} = c_1 \bigg(\dfrac{J}{J^\text{M}}\bigg)^2 + c_2 \bigg(\dfrac{J}{J^\text{M}}\bigg)
$$ 
where the rotational frequency $f_c$ and the angular momentum $J$ are normalised using respective properties ($f_c^\text{M}$ and $J^\text{M}$) for the maximum mass differentially rotating NS for each EoS from Table \ref{tab:max_mass}. The coefficients for the fitted polynomial for each EoS are listed in Table \ref{tab:quasi_fit}. We note that this fit also depends on the differential rotation profile adopted for the study, we will look at the effects of other differential rotation profiles in our future work. 

\iffalse
For the equilibrium sequences obtained, the stability criterion is given by
$$
\frac {\partial J}{\partial \rho_c}\bigg|_{M_b=\text{const}}<0
$$
The $\rho_c$ increases as the one moves down along the constant $M_b$ curves. An unstable configuration for a given $M_b$ is always at a higher frequency than the stable configuration having the same angular momentum. 
\fi
\begin{table}[!h]
\tabularfont
\centering
\begin{tabular}{ccc}
    \hline
         Configuration & $c_1$ & $c_2$ \\
        \hline     
         DD2, T=0 & -0.995 & 1.982 \\
         DD2, s=2 & -0.794 & 1.803 \\
         BHB$\Lambda\phi$, T=0 & -1.093 & 2.089 \\
         BHB$\Lambda\phi$, s=2 & -0.729 & 1.732 \\
        \hline 
    \end{tabular}
    \caption{The values of the fitting parameters for the polynomial function relating rotational frequency to angular momentum of differentially rotating ($a=0.2$) HMNS at the quasi-stationary stability limit.}
    \label{tab:quasi_fit}
\end{table}

\begin{table}[!h]
\tabularfont
\centering
\begin{tabular}{cccc}
    \hline
         Configuration & $M_b(M_{sol})$ & $\Delta f_c(Hz)$ & $\Delta$J($GM_{sol}^2/c$) \\
        \hline     
         DD2, T=0 & 3.11 & 79.58 & 0.45 \\
         DD2, s=2 & 3.11 & 43.61 & 0.33 \\
         \hline
         DD2, T=0 & 2.89 & 71.72 & 0.37 \\
         BHB$\Lambda\phi$, T=0 & 2.89 & 7.73 & 0.42 \\
        \hline 
    \end{tabular}
    \caption{Change in frequency and angular momentum in the transition from differential rotation ($a=0.2$) to uniform rotation (Red sequences plotted in Fig. \ref{Fig:Freq_comp}). The first two rows compare NS of the same composition and $M_b$ at different temperatures. The last two rows compare cold NS having the same $M_b$ but different compositions.}
    \label{tab_diff_unif}
\end{table}

In the presence of $\Lambda$ hyperons, the spin-up is more robust. If we follow the spin evolution of a cold NS of baryon mass $2.89M_{sol}$ with DD2 and BHB$\Lambda\phi$ EoSs, this becomes more evident. Here, the NS with hyperons spins up to $1530Hz$, whereas a NS with only nucleons, spins down from $1290Hz$. The spin-up rate against angular momentum is further enhanced by increasing entropy. For example a strange, hot NS with baryon mass $2.75M_{sol}$ spins up from $1280Hz$ to $1370Hz$ for a small angular momentum loss of 0.1 $GM^2_{sol}/c$. For its cold counterpart, the corresponding rise in frequency is $130Hz$, over a larger angular momentum loss of 1.07 $GM^2_{sol}/c$. A similar trend can be seen for the DD2 EoS at s=2.

We also investigate the role of differential rotation and confirm that a star with a higher degree of differential rotation can support a higher maximum mass.
In Fig. \ref{Fig:Freq_j_diffa}, the rotational frequency of  differentially rotating sequences for $a=0.1$ and $a=0.3$ is plotted as a function of angular momentum  for the DD2 EoS at T=0. The Kepler frequency corresponding to the maximum mass NS is $1600Hz$ for $a=0.1$, and $2100Hz$ for $a=0.3$. When two supramassive constant rest mass sequences  are compared for different degrees of differential rotation, the back-bending is more pronounced for the case with higher value of $a$.
%\textcolor{green}{\textit We may use the following paragraph in Result section}
%For this, we plot rest mass sequences for different central enthalpies in line with previous work done on uniformly rotating neutron stars [2]. The rest mass sequences are generated to simulate a neutron star transitioning from differential rotation to uniform rotation and static cases for constant baryon mass. Newly born neutron stars slowly transition from differential to uniform rotation, this damping in differential rotation can be attributed to magnetic braking and viscosity [3], which can be tackled in future work.

%Rotational Frequency vs T/W

In Fig. \ref{Fig:Freq_tw}, we can see the relationship between the frequency and the ratio of rotational kinetic energy to gravitational binding energy (T/W) for the same sets of cold and hot EoSs. The solid curves correspond to Kepler limits for NS rotating uniformly (orange) and differentially (blue).  We note that a differentially rotating star can reach a much larger T/W ratio compared to a uniformly rotating star. The
triaxial instability could play a more important role as the T/W value can be as high as 0.164 for a differentially rotating NS with the DD2 T=0 configuration, which comes down to 0.136 for the uniformly rotating one. Moreover, the T/W values are lesser for higher entropy stars than cold stars. We have seen that a hotter star is larger in size \citep{Neelam2018, apj2020}, or in other words, thermal pressure results in a less tightly bound star. This was also reported by
\citep{Camelio2021} for a different rotation law. It is further lowered for a softer EoS, in the presence of $\Lambda$ hyperons. Hence the NS with strange matter and/or higher entropy are stable with respect to axisymmetric perturbation. The maximum values of T/W for different cases considered here are listed in Table 1. The bar-modes
in a rapidly rotating, gravitationally bound fluid body may
be excited only for cold NS (T/W $\geq 0.14$). For general relativistic stars, the instability may occur at lower values of $\text{T/W} \sim 0.08$, with gravitational wave-emission \citep{Cook1994}. For all the cases, instability may occur before the NS can reach the mass-shedding limits. However, a stable configuration may exist with smaller T/W as the stars spin-up.
%A&A 559, A118 (2013) N. S. Ayvazyan, G. Colucci, D. H. Rischke, and A. Sedrakian
The constant rest mass sequences are plotted with dashed lines. Here, all the normal sequences  spin down as they can evolve from a differentially rotating state to a non-rotating/static state while being stable under quasi radial perturbations. One can observe the back-bending phenomenon in this graph, wherein, the constant $M_b$ curves for supramassive stars turn up to intersect the quasi stationary stability curve (dash-dotted) and eventually reach the Kepler curve, instead of hitting the origin like the normal sequences do. This was also observed by \citep{Cook1994} for uniform rotation. 
The qualitative features for the NS with nucleons-only or strange matter are more or less the same. Both the nucleonic and strange models at T=0
with higher masses may undergo an instability starting
at some frequency, but spin-up and attain stable configurations. 

%textcolor{orange}{(The supramassive sequences plotted don't go to static configuration)} 
Finally, we compare two evolution scenarios for NS with the same rest mass. In case 1, the star continues to rotate differentially throughout the evolution with no change in $a$. In case 2, the differential rotation gradually decays to uniform rotation as the star evolves. For the two cases, the constant rest mass sequences in the $M_g \text{ vs } \rho_c$ space overlap within systematic error. However, the back-bending patterns change with angular momentum and T/W loss. The constant rest mass ($M_b=3.11M_{sol}$) sequences for the two cases are plotted in Fig. \ref{Fig:Freq_comp} case the DD2 EoS with different entropies. At any given angular momentum, a case 1 star rotates at a higher frequency than a case 2 star. 

We tabulate the change in frequency and angular momentum for the transition from differential rotation ($a=0.2$) to uniform rotation for different profiles in Table. \ref{tab_diff_unif}. In the first two rows, we see that increasing the temperature for a fixed composition leads to a smaller loss in both frequency and angular momentum during the transition. Additionally, in the last two rows, we observe that the change in frequency is significantly smaller for a star which contains hyperons when compared to a star with a nucleonic EoS. On the other hand, the change in angular momentum is slightly larger for the star containing hyperons compared to the star with the nucleonic EoS.

%\begin{table}[htb]
%% use tabular font for a smaller size font
%\tabularfont
%\caption{Table fitting in a single column.}\label{tableExample} %%10/12
%\begin{tabular}{lccccc}
%\topline
%one& two &three&four&five&six\\\midline
%1&2&3&4&5&6\\
%aaa&bbbb& ccccc&dddd&eeeee&ffffff\\
%\hline
%\end{tabular}
%%use \tablenotes{footnote} to get the table foot note
%\tablenotes{Sample table footnote}%%9/11
%\end{table}

%%Use table* environment to get the table spanning both the columns

%\begin{table*}[htb]
%\tabularfont
%\caption{Caption text here}\label{secondTable}
%\begin{tabular}{lccccccccccccr}
%\topline
%\textbf{head1}&\multicolumn{11}{c}{\textbf{head2}}&\textbf{head3}\\
%\midline
%one& two &three&four&five&six&seven&eight&nine&ten&eleven&twelve&thirteen\\
%1&2&3&4&5&6&7&8&9&10&11&12&13\\
%aaa&bbbb&cccc&ddddd&eee&ffff&ggggg&hhhhhhhh&iiii&kkkkkk&hhh&jjjjjj&lllll\\
%\hline
%\end{tabular}
%\tablenotes{Table footnote here. Table spanning both the columns.}
%\end{table*}

%%An example of a figure

%\begin{figure}[!t]
%\includegraphics[width=.8\columnwidth]{fig1.eps}
%\caption{caption goes here}\label{figOne}
%\end{figure}

%%An example of a double column figure
%%Use figure* environment

%\begin{figure*}
%\centering\includegraphics[height=.15\textheight]{fig1.eps}
%\caption{caption spanning two columns}
%\centering\includegraphics[height=.25\textheight]{fig1.eps}
%\caption{caption here}
%\end{figure*}

\section{Summary and Discussions}\label{sec:sum}
We have generated constant rest mass sequences for isolated differentially rotating neutron stars and analysed the effect of strangeness and finite temperature on their equilibrium sequences. We have found that our results for the maximum mass and frequency vary significantly based on the choice of EoS and the degree of differential rotation. Additionally, we see that both increasing temperature and addition of hyperons tend to boost the spin-up in differentially rotating NS. While these general trends are more robust, we go into more detailed discussions of our study one by one. 

First, we report the effect of differential rotation on the maximal mass and rotational frequency of a NS. It is observed that differential rotation enables a NS to support a significantly higher mass than a non-rotating NS. This relative increase in maximum supported mass is higher in cold stars than hot stars. Differential rotation also allows the star to attain higher central rotational frequency than a uniformly rotating NS. Further increase in the degree of differential rotation enables the star to support even higher mass and rotational frequency while staying within the Kepler mass-shedding limit. For the j-constant law, the degree of differential rotation has no effect on a constant rest mass sequence being normal or supramassive. This classification depends on the mass of the NS and the choice of the EoS.

We also obtain constant rest mass sequences which do not cross the Kepler curve for uniform rotation (for example: $M_b$ = 3.20$M_{sol}$ curve for DD2 s=2 EoS and $M_b$ = 2.75$M_{sol}$ curve for BHB$\Lambda\phi$ EoS). These curves, theoretically, represent stars that cannot be stabilised by uniform rotation. Stars born with this configuration will be highly unstable, as reduction in the degree of differential rotation through processes such as magnetic braking and viscosity 
%\textcolor{orange}{(need to add more processes?)}
would inevitably lead to a collapse. This collapse will occur because a uniformly rotating counterpart to this configuration does not exist. This avenue is not well explored in our study and may be a subject for future work.

When investigating the spin-up phenomenon, we observe that the presence of hyperons in the EoS makes the spin-up more robust. The NSs following supramassive sequences with a cold nucleonic EoS have very high masses, which makes them unlikely candidates for observation. But, the presence of hyperons can make the EoS softer, which allows relatively lower mass stars to follow supramassive sequences. Moreover, we observe that a higher temperature can help boost the spin-up rate further. Therefore, a hot star with hyperons can be a viable candidate for modelling observations of spin-up in massive NS. 
%We have also fitted the quasi-stationary limit for equilibrium sequences using a polynomial function.

We find that the maximum T/W attained by the NS decreases with higher entropy. Thermal pressure or   the presence of strangeness is 
observed to increase the stability of the NS by reducing the maximal T/W values. However, high mass stars with nucleonic or strange models at T=0 may undergo an instability at some frequency to attain stable configurations upon spin-up. 

We also compare the supramassive sequences of a star that rotates differentially throughout its evolution and a star that transitions from differential rotation to uniform rotation. The latter case is more plausible as differential rotation can be damped by physical processes on a timescale which is shorter than the star's age. In the equilibrium sequences for a transition from differential to uniform rotation, the spin-up starts only after the transition to uniform rotation is complete. This means that a stable transition might require the frequency of the star to decrease with the degree of differential rotation, to keep the star within the Kepler mass-shedding limit. For two stars with the same baryon mass and composition, increasing the temperature decreases the rate of spin-down during the transition. When we compare two stars with the same baryon mass but different compositions, we see that the presence of hyperons drastically reduces the rate of spin down during the transition from differential to uniform rotation.

We find that differentially rotating hot stars with hyperons can be a viable option to model the HMNS observed as  remnants of binary NS mergers. Our results also provide evidence for the enhancement of spin-up in the presence of hyperons in the EoS. For our choice of EoS, spin-up is also supported by an increase in the temperature of the HMNS. This work opens up avenues for further investigation of differentially rotating HMNS with different rotation laws, compositions and temperatures. We plan to explore these avenues in detail in subsequent papers.  % config -> freq change | angumom change | ratio
% 3.11 mb dd2 s0 -> 1368.779 - 1289.203 | 5.5171 - 5.0620 | 173.5882
% 3.11 mb dd2 s2 -> 1268.761 - 1225.150 | 4.9280 - 4.6017 | 131.7806
% 2.89 mb dd2 s0 -> 1284.976 - 1213.260 | 4.8192 - 4.4525 | 193.6135
% 2.89 mb bhb s0 -> 1364.687 - 1356.957 | 4.6746 - 4.2507 |  18.8724

%%Use section* for acknowledgements
\section*{Acknowledgements}\label{sec:ack}
Authors are grateful to Micaela Oertel and Prasanta Char for insightful discussions.

\vspace{-1em}

%%use \balance somewhere in the left column of the last page to balance the two columns in the end page

%%References section
\bibliography{paper_spr}

\end{document}